\documentstyle[preprint,aps,eqsecnum]{revtex}
\tighten
\begin{document}
\preprint{PITT-96-377; CMU-HEP96-02; DOE/ER/40682-113; hep-ph/9603337}
\draft
\title{\bf THE GAUGE INVARIANT EFFECTIVE POTENTIAL:\\
EQUILIBRIUM AND NON-EQUILIBRIUM ASPECTS}
\author{{\bf D. Boyanovsky$^{(a)}$, 
D. Brahm$^{(b)}$   R. Holman$^{(b)}$ and 
D.-S. Lee$^{(c)}$}}   
\address
{(a) Department of Physics and Astronomy, University of
Pittsburgh, Pittsburgh, PA. 15260, U.S.A. \\
 (b) Department of Physics, Carnegie Mellon University, Pittsburgh,
PA. 15213, U. S. A. \\
 (c) Department of Physics and Astronomy, University of North Carolina\\
Chapel Hill, N.C. 27599, U.S.A. }
\date{}
\maketitle
\begin{abstract}
We propose a gauge invariant formulation of the effective potential in terms
 of
a gauge invariant order parameter, for the Abelian Higgs model. The one-loop
contribution at zero and finite temperature is computed explicitly, and the
leading terms in the high temperature expansion are obtained. The result is
contrasted to the effective potential obtained in several covariant
gauge-fixing schemes, and the gauge invariant quantities that can be reliably
extracted from these are identified. It is pointed out that the gauge invariant
effective potential in the one-loop approximation is complex for {\em all
values} of the order parameter between the maximum and the minimum of the tree
level potential, both at zero and non-zero temperature. The imaginary part is
related to long-wavelength instabilities towards phase separation. We study the
real-time dynamics of initial states in the spinodal region, and relate the
imaginary part of the effective potential to the growth rate of equal-time
gauge invariant correlation functions in these states. We conjecture that the
spinodal instabilities may play a role in non-equilibrium processes {\em
inside} the nucleating bubbles if the transition is first order.
\end{abstract}
\pacs{11.15.Ex; 11.10.-z} 

\section{\bf Introduction and Motivation}

In this article we are concerned with the effective potential in gauge
theories. It has been recognized very early on, that the effective potential is
a gauge dependent quantity\cite{dolan} and only a limited amount of information
extracted from it is actually physically meaningful. This gauge dependence can
be understood from several equivalent points of view. The effective potential
can be identified with the generating functional of 1-particle irreducible
Green's functions (the effective action) at zero four momentum transfer, and
therefore it is an off-shell quantity. Alternatively, the effective potential
is identified with the energy (or free energy) of a particular state (or
ensemble) constrained to have an homogeneous expectation value of the scalar
field.

The energy, or the free energy, is usually calculated by fixing a particular
gauge in the path-integral. In a gauge theory, the (complex) scalar fields
transform under gauge transformations and their expectation value in a {\em
gauge fixed state or ensemble} is obviously a gauge dependent quantity.
Despite this shortcoming it has been recognized that certain quantities are
gauge independent. Dolan and Jackiw\cite{dolan} recognized that the critical
temperature is a gauge invariant quantity and recently Metaxas and
Weinberg\cite{meta} used the Nielsen identities\cite{nielsen} to prove that the
bubble nucleation rate {\em at zero temperature} is gauge invariant (we are not
aware of a similar proof at finite temperature). The gauge invariance of these
quantities can be understood from the fact that they are associated with
homogeneous and inhomogeneous extrema of the effective action respectively;
these are known to be gauge invariant.

Considerable effort has been devoted to constructing a gauge invariant
effective action and effective potential\cite{vilko}, of which the background
field method of Vilkovisky and Dewitt\cite{vilko} is the most
popular\cite{rebhan}. These formulations of effective actions are technically
formidable and do not readily lend themselves to a manageable formulation of
equilibrium or non-equilibrium descriptions. Furthermore, it has recently been
pointed out using the pinch-technique, that despite its formal gauge
invariance, implementation of the background field method requires a gauge
fixing parameter for the {\em fluctuations}. This leads to a gauge-parameter
dependence in finite parts of self-energies at finite
temperature\cite{pinch,sasaki}, which in turn leads to a gauge dependence of
the thermal renormalization group beta function as discussed in detail by
Sasaki\cite{sasaki}.

Alternative formulations of effective potentials have been offered in terms of
a radial and angular decomposition of the complex scalar
fields\cite{lawrie,tye} or alternatively in terms of gauge invariant composite
operators\cite{fodor}.  There are several shortcomings in the formulation of
the effective potential in terms of the radial field variable\cite{lawrie,tye}
or composite operators\cite{fodor}. This variable is understood as the ``square
root'' of a composite operator that requires a {\em subtraction} to be
renormalized; shortcomings of this approach had been already
recognized\cite{lawrie}.  Furthermore in the path integral evaluation there is
an ambiguous Jacobian arising from the change of variables to radial and
angular fields. This Jacobian has to be incorporated in the perturbative
expansion to obtain a consistently renormalized effective action\cite{lawrie}.

However, even when these technicalities are overcome by some renormalization
scheme (such as dimensional regularization), it is conceptually unclear how to
interpret symmetry breaking in terms of the radial field. At the operator
level, the radial field variable acquires a ground state expectation value {\em
even in the symmetric phase} as can be seen with a simple example of a
two-dimensional isotropic harmonic oscillator. Composite bilinear operators
typically require subtractions to be renormalized, and their expectation value
is therefore ambiguous.

Our motivation is to obtain a gauge invariant description of the effective
potential (and eventually the effective action) and to use it to provide some
preliminary information on the dynamics of non-equilibrium processes during the
phase transition in gauge theories. In view of the above discussion and
critique of previous approaches, such an enterprise is clearly worthwhile
because only a truly gauge invariant description of effective potentials can be
considered trustworthy in terms of extracting physical quantities such as
supercooling temperature, latent heat and others that are very important in the
quantitative description of non-equilibrium features.

Our program for the construction of an effective potential can be summarized in
the following steps: a) select the gauge invariant states of the theory,
namely, those that are annihilated by the first class constraints (such a
selection {\em will not} involve any gauge fixing), b) recognize a gauge
invariant order parameter that is invariant under local gauge transformations,
but transforms non-trivially under the global symmetry that can be
spontaneously broken, and c) construct the effective potential for this gauge
invariant order parameter. An attempt to establish a finite temperature
framework in terms of gauge invariant states has been reported
previously\cite{herby} within a different context and with a different goal,
but to our knowledge it has not been implemented or attempted within the
context of the effective potential.

In this article we focus on such a description for the Abelian Higgs model
(scalar electrodynamics) and we expect to generalize the procedure and its
quantitative implementation to Yang-Mills theories in the near future.

In section II we implement the first step of the program, that is, we select
the gauge invariant states and order parameter {\em without} fixing a gauge, by
requiring that the physical states be annihilated by the first class
constraints of the theory which are recognized as the generators of local gauge
transformations. Gauge invariant operators are then recognized as those that
commute with these constraints, out of which we recognize the proper order
parameter. In section III we explicitly construct the one loop effective
potential both at zero and non-zero temperature and compare our results with
those obtained in popular covariant gauges. From this comparison we establish
when the gauge-fixed results lead to physical (gauge independent)
predictions. In this section we also argue that the gauge dependence of the
usual (gauge fixed) effective potential is not relieved by hard-thermal loop
resummation.  We also provide the high-temperature expansion of the gauge
invariant effective potential and point out that the ``cubic'' terms which are
typically taken as a signal of the strength of a first order transition are in
general complex and gauge dependent in fixed gauge path integral calculations
of the effective potential.

In section IV we use the gauge invariant effective potential to study the early
time behavior of spinodal phase separation and the instabilities associated
with the spinodal line in gauge theories. We establish a correspondence between
the imaginary part of the one loop gauge invariant effective potential and the
rate of growth of correlations in the spinodal region. Two appendices are
devoted to some technical details.

\section{\bf The Gauge Invariant Description:}

The focus of our study is scalar electrodynamics or the Abelian Higgs model
whose Lagrangian density is
\begin{eqnarray}
 {\cal{L}} & = & -\frac{1}{4}F^{\mu \nu} F_{\mu \nu}+D^{\mu}\phi^{\dagger}
 D_{\mu}\phi- \lambda (\phi^{\dagger}\phi-\mu^2)^2 \label{higgslag}\\
 D_{\mu}\phi & = & (\partial_{\mu} \phi +i e A_{\mu} \phi) \label{covder}
\end{eqnarray}

The description in terms of gauge invariant states and operators is best
achieved within the canonical formulation, which begins with the identification
of canonical field variables and constraints. These will determine the
classical physical phase space and, at the quantum level, the physical Hilbert
space.

The canonical momenta conjugate to the scalar and vector fields are given by
\begin{eqnarray}
\Pi^0 & = & 0 \label{pioo} \\ \Pi^i & = & \dot{A}^i+\nabla^i A^0 = -E^i
\label{amom}\\ \pi^{\dagger} & = & \dot{\phi}+i e A^0 \phi \label{phidagmom}\\
\pi & = & \dot{\phi}^{\dagger}-i e A^0 \phi^{\dagger} \label{phimom}
\end{eqnarray}

The Hamiltonian is therefore
\begin{eqnarray}
H= && \int d^3x \left\{ \frac{1}{2}\vec{\Pi}\cdot
\vec{\Pi}+\pi^{\dagger}\pi+(\vec{\nabla}\phi- ie \vec{A}\phi)\cdot
(\vec{\nabla}\phi^{\dagger}+ie \vec{A}\phi^{\dagger})+\frac{1}{2} (\vec{\nabla}
\times \vec{A})^2 + \right. \nonumber \\ &&\left. \lambda
(\phi^{\dagger}\phi-\mu^2)^2 + A_0\left[\vec{\nabla}\cdot \vec{\Pi} - ie (\pi
\phi-\pi^{\dagger}\phi^{\dagger})\right] \right\} \label{hamiltonian}
\end{eqnarray}

There are several different manners of quantizing a gauge theory, but the one
that exhibits the gauge invariant states and operators, originally due to
Dirac, begins by recognizing the first class constraints (mutually vanishing
Poisson brackets). From here there are several possibilities: i) the
constraints become operators in the quantum theory and are imposed onto the
physical states, thus defining the physical subspace of the Hilbert space and
gauge invariant operators. ii) Introduce a gauge, converting the first class
system of constraints into a second class (with non-zero Poisson brackets
between the constraints) and introducing Dirac brackets.  This is the popular
way of dealing with the constraints and leads to the usual gauge-fixed path
integral representation\cite{slavnov} in terms of Faddeev-Popov determinants
and ghosts.

We will instead proceed with the first possibility that leads to an unambiguous
projection of the physical states and operators. Such a method has been
previously used by James and Landshoff within a different
context\cite{landshoff}.

In Dirac's method of quantization\cite{hat} there are two first class
constraints which are:
\begin{equation} 
\Pi^0= \frac{\delta {\cal{L}}}{\delta A^0} = 0 \label{pi0const}
\end{equation}
and Gauss' law:
\begin{eqnarray}
{\cal{G}}(\vec{x},t) & = & \nabla^i \pi^i - \rho=0 \label{gausslaw} \\ \rho & =
& i e \left(\phi \pi- \phi^{\dagger}\pi^{\dagger}\right) \label{rho}
\end{eqnarray}
with $\rho$ being the matter (complex scalar) field charge density.

Gauss' law can be seen to be a constraint in two ways: either because it cannot
be obtained as a Hamiltonian equation of motion, or because in Dirac's
formalism, it is the secondary (first class) constraint obtained by requiring
that the primary constraint (\ref{pi0const}) remain constant in
time. Quantization is now achieved by imposing the canonical equal-time
commutation relations
\begin{eqnarray}
\left[\pi^0(\vec{x},t),A^0(\vec{y},t)\right] & = & -i \delta(\vec{x}-\vec{y})
\label{a0comrel} \\ \left[\pi^i(\vec{x},t),A^j(\vec{y},t)\right] & = & -i
\delta^{ij} \delta(\vec{x}-\vec{y}) \label{acomrel}\\
\left[\pi^{\dagger}(\vec{x},t),\phi^{\dagger}(\vec{y},t)\right] & = & -i
\delta(\vec{x}-\vec{y})
\label{fidaggercom}\\
\left[\pi(\vec{x},t),\phi(\vec{y},t)\right] & = & -i \delta(\vec{x}-\vec{y})
\label{ficom}
\end{eqnarray}

In Dirac's formulation, the projection onto the gauge invariant subspace of the
full Hilbert space is achieved by imposing the first class constraints onto the
states. Physical operators are those that commute with the first class
constraints. With the above equal-time commutation relations it is
straightforward to see that the unitary operator
\begin{equation}
U_{\Lambda}= \exp{\left\{i \int \left[\Pi^0 \dot{\Lambda}+{\cal{G}} \Lambda
\right] d^3x
\right\}} \label{gaugeop}
\end{equation} 
performs the local gauge transformations. Thus the first class constraints are
recognized as the generators of gauge transformations. In particular, Gauss'
law (\ref{gausslaw}) is the generator of time independent gauge
transformations. Requiring that the physical states be annihilated by these
constraints is tantamount to selecting the gauge invariant states.
Consequently operators that commute with the first class constraints are gauge
invariant.

In the Schroedinger representation, in terms of wave-functionals, the canonical
momenta are represented by hermitian differential operators, and the
constraints applied onto the states become functional differential equations
that the wave-functionals must satisfy:
\begin{eqnarray}
&& \frac{\delta}{\delta A_0(\vec{x})} \Psi[A,\phi,\phi^{\dagger}] = 0
 \label{1stconst} \\ && \left[ \vec{\nabla}_x \frac{\delta}{\delta
 \vec{A}(\vec{x})}-i e \left(\phi(\vec{x})\frac{\delta}{\delta \phi(\vec{x})}-
 \phi^{\dagger}(\vec{x})\frac{\delta}{\delta \phi^{\dagger}(\vec{x})}\right)
 \right] \Psi[A,\phi,\phi^{\dagger}] = 0 \label{2ndconst}
\end{eqnarray} 
The first equation simply means that the Schroedinger wave-functional does not
depend on $A_0$, whereas the second equation means that the wave-functional is
only a functional of the combination of fields that is annihilated by the
Gauss' law functional differential operator.  It is a simple calculation to
prove that the fields
\begin{eqnarray}
\Phi(\vec{x}) & = & \phi(\vec{x})\exp\left[ie \int d^3 y \vec{A}(\vec{y})\cdot
\vec{\nabla}_y G(\vec{y}-\vec{x})\right]\label{gauginvphi} \\
\Phi^{\dagger}(\vec{x}) & = & \phi^{\dagger}(\vec{x})\exp\left[-ie \int d^3 y
\vec{A}(\vec{y})\cdot \vec{\nabla}_y G(\vec{y}-\vec{x})\right]
\label{gauginvphidag}
\end{eqnarray}
are annihilated by Gauss' law functional differential equation with $
G(\vec{y}-\vec{x})$ the Coulomb Green's function that satisfies
\begin{equation}
\nabla^2 G(\vec{y}-\vec{x})=0 \label{coulombgf}
\end{equation}
Furthermore writing the gauge field into transverse and longitudinal
components as follows
\begin{eqnarray}
&& \vec{A}(\vec{x}) = \vec{A}_L(\vec{x})+\vec{A}_T(\vec{x}) \label{fieldsplit}
\\ && \vec{\nabla} \times \vec{A}_L(\vec{x}) = 0 \label{long} \\ &&
\vec{\nabla} \cdot \vec{A}_T(\vec{x}) = 0 \label{trans}
\end{eqnarray}
it is clear that
\begin{equation}
\vec{\nabla}_x \frac{\delta}{\delta \vec{A}(\vec{x})} = \vec{\nabla}_x
\frac{\delta}{\delta \vec{A}_L(\vec{x})}. \label{difflong}
\end{equation}
Therefore the ``transverse component'' $\vec{A}_T(\vec{x})$ is also annihilated
by the Gauss' law operator. This analysis shows that the wave-functional
solutions of the functional differential equations that represent the
constraints in the Schroedinger representation are of the form
\begin{equation}
\Psi[\vec{A},\phi,\phi^{\dagger}] = \Psi[\vec{A}_T,\Phi,\Phi^{\dagger}]
\label{gaugeinvfunc}
\end{equation}
The fields $\vec{A}_T \; ; \Phi \; ; \Phi^{\dagger}$ are {\em gauge invariant}
as they commute with the constraints. The canonical momenta conjugate to $\Phi
\ \Phi^{\dagger}$ are found to be
\begin{eqnarray}
\Pi(\vec{x}) & = & \pi(\vec{x}) \exp\left[-ie \int d^3 y \vec{A}(\vec{y})\cdot
\vec{\nabla}_y G(\vec{y}-\vec{x})\right] \label{canophi} \\
\Pi^{\dagger}(\vec{x}) & = & \pi^{\dagger}(\vec{x}) \exp\left[ie \int d^3 y
\vec{A}(\vec{y})\cdot \vec{\nabla}_y G(\vec{y}-\vec{x})\right]
\label{canophidag}
\end{eqnarray}
The momentum canonical to $\vec{A}\; ; \; \vec{\Pi}$ is written in terms of
``longitudinal'' and ``transverse'' components
\begin{equation}
\vec{\Pi}(\vec{x}) = \vec{\Pi}_l(\vec{x}) +\vec{\Pi}_T(\vec{x}) \label{PiA}
\end{equation}
both components are gauge invariant.

In the physical subspace of gauge invariant wave-functionals, matrix elements
of $\vec{\nabla}\cdot \vec{\Pi}$ can be replaced by matrix elements of the
charge density $\rho$ . Therefore in all matrix elements between gauge
invariant states (or functionals) one can replace
\begin{equation}
\vec{\Pi}_L(\vec{x}) \rightarrow ie \vec{\nabla}_x \int d^3y G(\vec{x}-\vec{y})
\left(\Phi \Pi - \Phi^{\dagger}\Pi^{\dagger}\right)(\vec{y}) \label{coulomb}
\end{equation}

Finally in the gauge invariant subspace the Hamiltonian becomes
\begin{eqnarray}
H= && \int d^3x \left\{ \frac{1}{2}\vec{\Pi}_T\cdot
\vec{\Pi}_T+\Pi^{\dagger}\Pi+(\vec{\nabla}\Phi- ie \vec{A}_T \Phi)\cdot
(\vec{\nabla}\Phi^{\dagger}+ ie \vec{A}_T \Phi^{\dagger})+\frac{1}{2}
(\vec{\nabla} \times \vec{A}_T)^2 + \right. \nonumber \\ && \left. \lambda
(\Phi^{\dagger}\Phi-\mu^2)^2 \right\} + \frac{1}{2}\int d^3y \int d^3x
\rho(\vec{x})G(\vec{x}-\vec{y})\rho(y)
\label{gauginham}
\end{eqnarray}
Clearly the Hamiltonian is gauge invariant, and it manifestly has the global
$U(1)$ gauge symmetry under which $\Phi$ transforms with a constant phase, 
$\Pi$ transforms with the opposite phase and
$\vec{A}_T$ is invariant.

This Hamiltonian is reminiscent of the Coulomb gauge Hamiltonian, but we
emphasize that we have not imposed any gauge fixing condition. The formulation
is fully gauge invariant, written in terms of operators that commute with the
generators of gauge transformations and states that are invariant under these
transformations.

There is a definite advantage in this gauge invariant formulation: the
(composite) field $\Phi(\vec{x})$ is a candidate for a {\em locally gauge
invariant order parameter}. The point to stress is the following: this operator
is {\em invariant} under local gauge transformations generated by the unitary
transformation $U_{\Lambda}$ given by equation (\ref{gaugeop}), that is
\begin{equation}
U_{\Lambda}\Phi(\vec{x})U^{-1}_{\Lambda}= \Phi(\vec{x})
\end{equation}
whereas it transforms as a charged operator under the {\em global} gauge
transformations generated by $Q=\int d^3x \rho(\vec{x})$, that is
\begin{equation}
e^{i\alpha Q}\Phi(\vec{x}) e^{-i\alpha Q}= e^{ie\alpha} \Phi(\vec{x}).
\end{equation}

Because the gauge constraints annihilate the physical states and these
constraints are the generators of local gauge transformations, these states are
invariant under the local gauge transformations and any operator that {\em is
not} invariant under these local transformations {\em must} have zero
expectation value. The {\em local} gauge symmetry cannot be spontaneously
broken; this result is widely known in lattice gauge theory as Elitzur's
theorem\cite{elitzur}. However, the {\em global} symmetry generated by the
charge $Q$ {\em can} be spontaneously broken and the expectation value of a
charged field signals this breakdown.

{}From this discussion we clearly see that a trustworthy order parameter must
be invariant under the local gauge transformations, thus commuting with the
gauge constraints, but must transform non-trivially under the global gauge
transformation generated by the charge.  The field $\Phi$ fulfills these
criteria and is the natural candidate for an order parameter.

\section{\bf The Effective Potential}

\subsection{\bf Zero Temperature}

We are now in condition to define the gauge invariant effective
potential. Consider the gauge invariant state $|\Psi;\chi \rangle$ such that
the expectation value of the gauge invariant order parameter $\Phi(\vec{x})$ in
this state is nonzero and space-time constant
\begin{equation}
\frac{\langle \Psi;\chi| \Phi(\vec{x})|\Psi;\chi\rangle}{\langle \Psi;\chi
|\Psi;\chi \rangle}= \chi \label{expval}
\end{equation}
The effective potential is defined as the minimum of the expectation value of
the Hamiltonian density in this state, namely
\begin{equation}
V_{eff}(\chi) = \frac{1}{\Omega} \mbox{min}\left\{ \frac{\langle \Psi;\chi| H
|\Psi;\chi\rangle}{\langle \Psi;\chi |\Psi;\chi \rangle} \right\}
\label{gauginveff}
\end{equation}
with $H$ being the gauge invariant, Hamiltonian given by equation
(\ref{gauginham}) and $\Omega$ the spatial volume\cite{suranyi}. The state
$|\Psi;\chi \rangle$ is chosen to minimize the expectation value of the
Hamiltonian subject to the constraint that the expectation value of $\Phi$ in
this state is $\chi$.

It is convenient to separate the expectation value of $\Phi$ as
\begin{equation}
\Phi(\vec{x}) = \chi + \eta(\vec{x}) \label{separ}
\end{equation}
The one-loop correction (formally of ${\cal{O}}(\hbar)$) to the effective
potential is obtained by keeping the {\em quadratic} terms in the Hamiltonian.
\begin{eqnarray}
H_q & = & \Omega \lambda (|\chi|^2- \mu^2)^2 + \int d^3x \left\{ \frac{1}{2}
\vec{\Pi}^2_T+ \frac{1}{2} (\vec{\nabla} \times \vec{A})^2 + e^2 \vec{A}^2_T
|\chi|^2 + \right.  \nonumber \\ & & \left. \Pi^{\dagger}\Pi +
(\vec{\nabla}\eta) (\vec{\nabla}\eta^{\dagger}) + 2 \lambda \eta^{\dagger} \eta
(|\chi|^2-\mu^2)+ (\eta \chi^{\dagger}+\eta^{\dagger}\chi)^2 \right\} +
\nonumber \\ & & \frac{e^2}{2} \int d^3 x d^3 y \left[\Pi(\vec{y})
\chi-\Pi^{\dagger}(\vec{y}) \chi^{\dagger}\right] G(\vec{y}-\vec{x})
\left[\Pi(\vec{x}) \chi-\Pi^{\dagger}(\vec{x}) \chi^{\dagger}\right]
\label{quadham}
\end{eqnarray}
The transverse components $\vec{A}_T$ describe a field with mass $m_T^2=
2e^2|\chi|^2$ and only two polarizations. The phase of $\chi$ can be absorbed
in $\Pi$ by a {\em global} phase transformation under which the Hamiltonian is
invariant.

It proves convenient to introduce real fields and canonical momenta as follows
\begin{eqnarray}
\eta & = & \frac{1}{\sqrt{2}}(\eta_1 + i \eta_2) \label{realfields} \\ \Pi & =
& \frac{1}{\sqrt{2}}(\Pi_1 - i \Pi_2) \label{realmomenta}
\end{eqnarray}
with $\{\Pi_{1,2}, \eta_{1,2}\}$ being independent canonical pairs. The
non-local part of the Hamiltonian is best treated in terms of the Fourier
transform of the fields and their canonical momenta, in terms of which the
quadratic part of the Hamiltonian finally becomes

\begin{eqnarray}
H_q & = & \Omega V_{cl}(|\chi|)+ \frac{1}{2}\sum_{k}\left\{\vec{\Pi}_T(k) \cdot
\vec{\Pi}_T(-k)+ \omega^2_T(k) \vec{A}_T(k) \cdot \vec{A}_T(-k) +
\right. \nonumber \\ & & \left. \Pi_1(k) \Pi_1(-k)+ \omega^2_H(k) \eta_1(k)
\eta_1(-k) + \Pi_2(k) \Pi_2(-k) \frac{\omega^2_T(k)}{k^2}+ \eta_2(k) \eta_2(-k)
\omega^2_g (k) \right\} \label{quadhamk}
\end{eqnarray}
where the frequencies are given in terms of the effective masses as
\begin{eqnarray}
\omega^2_T(k) = k^2+m^2_T &\;\;;\;\;& m^2_T = 2e^2|\chi|^2
  \label{transfreq} \\
\omega^2_H(k) = k^2+m^2_H &\;\;;\;\;& m^2_H = 2 \lambda (3|\chi|^2-\mu^2)
  \label{realfreq} \\
\omega^2_g(k) = k^2+m^2_g &\;\;;\;\;& m^2_g = 2 \lambda (|\chi|^2-\mu^2).
  \label{imfreq}
\end{eqnarray}
The last two terms can be brought to a canonical form by a Bogoliubov
transformation. Define the new canonical coordinate $Q$ and conjugate momentum
$P$ as
\begin{eqnarray}
\Pi_2(k) & = & \frac{k}{\omega_T(k)} P(k) \label{newmom} \\ \eta_2(k) & = &
\frac{\omega_T(k)}{k} Q(k) \label{newcoord}
\end{eqnarray}
in terms of which the last term of the Hamiltonian (\ref{quadhamk}) becomes a
canonical quadratic form with the {\em plasma} frequency
\begin{equation}
\omega^2_p(k) = \omega^2_g(k) \frac{\omega^2_T(k)}{k^2}=
 [k^2+2\lambda(|\chi|^2-\mu^2)][k^2+2e^2 |\chi|^2]/k^2 \label{plasmafreq}
\end{equation}

There are four physical degrees of freedom. The modes with frequency
$\omega_T(k)$ are the two transverse degrees of freedom, the mode with
frequency $\omega_H(k)$ is identified with the Higgs mode. In absence of
electromagnetic interactions ($e=0$) the mode with frequency $\omega_p(k)$
represents the Goldstone mode whereas {\em in equilibrium}, namely at the
minimum of the tree level potential, when $|\chi| = \mu$, it represents the
plasma mode which is identified as the screened Coulomb interaction, and the
transverse and plasma modes all share the same mass. However, when the
expectation value of the order parameter acquires a non-equilibrium value,
away from the minimum of the tree level potential (at this order), this
collective mode does not describe a particle with a Lorentz covariant
dispersion relation. The frequency clearly shows the combination of the
Goldstone dispersion relation and the long-range Coulomb interaction typical of
a description in terms of the dynamical degrees of freedom.

The lack of manifest Lorentz covariance in the dispersion relation can be
understood as follows.  Although the complex field $\phi$ in the original
Lagrangian density is a Lorentz scalar, the gauge invariant combination $\Phi$
given by eq. (\ref{gauginvphi}) is not, although it is a rotational scalar. A
particular Lorentz frame has already been chosen in making the transverse and
longitudinal decomposition of the vector potential (\ref{fieldsplit}). The
state of lowest available energy is expected to be Lorentz invariant, but for
arbitrary $\chi$ there is a constraint in the space of functions and these
constrained states are not the lowest energy states in the functional space.
In a scalar theory these states are manifestly Lorentz invariant, simply
because all fields are Lorentz scalars.  With vector fields the situation is
more complicated and the constrained, gauge invariant states are in general not
manifestly Lorentz invariant.  However the lowest energy equilibrium states (at
this order corresponding to $\chi = \mu$) {\em are} manifestly Lorentz
invariant. We will see in detail in section IV that for $\chi \neq \mu$ these
states are not stationary states of the Hamiltonian, therefore the lack of
Lorentz covariance for these states is reconciled with their non-equilibrium
evolution.

The quadratic Hamiltonian is now diagonalized in terms of creation and
destruction operators for the quanta of each harmonic oscillator. The ground
state is the vacuum for each oscillator and is the state of lowest
energy. Therefore the one loop (${\cal{O}}(\hbar)$) contribution to the
effective potential is obtained from the zero point energy of the
oscillators. Therefore accounting for the two polarizations of the transverse
components we find:
\begin{equation}
V_{eff}(|\chi|) = V_{cl}(|\chi|)+ \frac{1}{2} \int \frac{d^3k}{(2\pi^3)}[2
\omega_T(k)+ \omega_H(k)+\omega_p(k)] \label{gauginveffpot}
\end{equation}

The normalized wave-functional that satisfies (\ref{expval}) and gives the
minimum expectation value of the Hamiltonian, thus determining effective
potential via (\ref{gauginveff}) is given by
\begin{eqnarray}
\Psi[A_T,\Phi^{\dagger},\Phi] & = & N \exp\left\{-\frac{1}{2}\int d^3x \int
d^3y \vec{A}_T(\vec{x}) \cdot \vec{A}_T(\vec{y}) K_T(\vec{x}-\vec{y}) \right\}
\times \nonumber \\ & & \exp\left \{ - \frac{1}{2} \int d^3x \int d^3y
\eta_1(\vec{x}) \eta_1(\vec{y}) K_H(\vec{x}-\vec{y}) \right\} \times \nonumber
\\ & & \exp\left \{ - \frac{1}{2} \int d^3x \int d^3y \eta_2(\vec{x})
\eta_2(\vec{y}) K_p(\vec{x}-\vec{y}) \right\} \label{wavefunctional} \\ N & = &
\Pi_{k}\left[\frac{\omega^2_T(k) \omega_H(k)
\omega_p(k)}{\pi^4}\right]^{\frac{1}{4}} \label{normalization} \\
K_T(\vec{x}-\vec{y}) & = & \int \frac{d^3k}{(2\pi)^3} \omega_T(k) e^{i \vec{k}
\cdot (\vec{x}-\vec{y})} \label{transkern} \\ K_H(\vec{x}-\vec{y}) & = & \int
\frac{d^3k}{(2\pi)^3} \omega_H(k) e^{i \vec{k} \cdot (\vec{x}-\vec{y})}
\label{higgskern} \\ K_p (\vec{x}-\vec{y}) & = & \int \frac{d^3k}{(2\pi)^3}
\frac{k^2 \omega_p(k) }{\omega^2_T(k)} e^{i \vec{k} \cdot (\vec{x}-\vec{y})}
\label{plaskern}
\end{eqnarray}
This Gaussian wave-functional is clearly gauge invariant, and it has the
correct limits: for $e=0$ ($\omega_T = k \; ; \omega_p=\omega_g$) gives the
(gauge invariant) wave-functional of free electromagnetism times the Gaussian
wave-functional of a complex scalar with the $U(1)$ global symmetry
spontaneously broken, which for $|\chi|^2 = \mu^2$ corresponds to the Higgs and
a Goldstone mode.  Writing the fluctuation fields $\eta_1 \; ; \; \eta_2$ in
terms of $\Phi \; ; \Phi^{\dagger}$, we clearly see that this wave functional
describes a broken symmetry state since under the global U(1) transformation
the wavefunctional is changed into an orthogonal wavefunctional in the infinite
volume limit.

The unbroken phase, with $\chi =0 \; ; -\mu^2 = m^2 >0$ corresponds to the
ground state wave-functional for free electromagnetism times the ground-state
wave-functional of two free real scalar fields with equal mass $m
\sqrt{2\lambda}$. It is straightforward to see that the expectation value of
the radial variable $\rho = \sqrt{\Phi^{\dagger}\Phi}$ is different from zero
in this phase and cannot be used as an order parameter to signal spontaneous
global symmetry breaking as discussed previously.

The $k-$ integrals in the effective potential (\ref{gauginveffpot}) are
performed with an upper momentum cutoff $\Lambda$. Neglecting a
$\chi$-independent term proportional to $\Lambda^4$ as well as terms that
vanish in the $\Lambda \rightarrow \infty$ limit, and introducing a
renormalization scale $\kappa$ we obtain the following (unrenormalized)
expression
\begin{eqnarray}
V_{eff}(|\chi|) &=& V_{cl}(|\chi|)+ \frac{1}{4\pi^2}\left\{
  \frac{\Lambda^2}{4}\left[3m^2_T+m^2_H+m^2_g \right] \right. \nonumber\\
&+& \left. \frac{1}{16}\ln\left(\frac{\kappa^2}{4\Lambda^2}\right) \left[
  3m^4_T+m^4_H+m^4_g-2m^2_Tm^2_g \right] \right. \nonumber\\
&+& \left. \frac{1}{16} \left[ 2m^4_T \ln(\frac{m^2_T}{\kappa^2})+ m^4_H
  \ln(\frac{m^2_H}{\kappa^2})+ (m^2_g-m^2_T)^2 \ln\left(\frac{m^2_T+m^2_g+
  2\sqrt{m^2_g m^2_T}}{\kappa^2}\right) \right] \right. \nonumber\\
&+& \left. \frac{1}{32}\left[3m^4_T+m^4_H+m^4_g+6m^2_g m^2_T\right] \right\}
\label{unrenoveff}
\end{eqnarray} 

The cutoff dependent terms can be absorbed in a renormalization of $\mu^2$
(terms of ${\cal{O}}(\Lambda^2)$ and ${\cal{O}}(\ln(\Lambda))$ proportional to
$|\chi|^2$) and the quartic coupling $\lambda$ (terms of
${\cal{O}}(\ln(\Lambda))$ proportional to $|\chi|^4$). Using this
renormalization prescription we find the following result for the renormalized
and gauge invariant one-loop effective potential
\begin{eqnarray} \lefteqn{
V_{eff,R}(|\chi|) = \lambda(|\chi|^2-\mu^2)^2+ \frac{1}{4\pi^2} \left\{
  \vphantom{\frac{\sqrt{m_g^2}}{\kappa^2}} 
  \frac{1}{32}\left[3m^4_T+m^4_H+m^4_g+6m^2_g m^2_T\right] \right.} \nonumber\\
 && \left.\qquad +\frac{1}{16} \left[ 2m^4_T \ln(\frac{m^2_T}{\kappa^2})+ m^4_H
  \ln(\frac{m^2_H}{\kappa^2})+ (m^2_g-m^2_T)^2 \ln\left(\frac{m^2_T+m^2_g+
  2\sqrt{m^2_g m^2_T}}{\kappa^2}\right) \right] \right\} \label{veffren}
\end{eqnarray}

In this expression $\lambda$ and $\mu$ and all masses are renormalized with
the above prescription).

In the region $|\chi|^2 < \mu^2/3$ the Higgs ``mass'' is purely imaginary ,
whereas for $|\chi|^2 < \mu^2, \ m^2_g <0$. Therefore we see that the
logarithmic contributions to the effective potential from Higgs and plasma
modes are {\em imaginary} (the last two logarithms in eq. \ref{veffren}),
whereas the contribution of the gauge boson is real. The region in which the
effective potential is imaginary is a region of unstable
states\cite{weinberg,danvega}, and the imaginary part of the effective
potential disguises a non-equilibrium situation whose dynamics will be
addressed in section IV. This region of instabilities for homogeneous
configurations is known as the spinodal region. In this region the system is
unstable to phase separation, and the imaginary part of the effective potential
appears as a result of attempting to describe an intrinsically time dependent
state as a stationary state via analytic continuation.

\subsection{Finite Temperature:}

The finite temperature effective potential is identified with the free energy
density under the constraint that the ensemble average of the field is given by
a space-time independent configuration $\chi$. That is
\begin{eqnarray}
V_T(\chi) & = & -T\ln({\rm Tr}\, \hat{\rho}) \label{finiteteffpot} \\
  \chi & = & \frac{{\rm Tr}\, \Phi(\vec{x})\hat{\rho}}{{\rm Tr}\, \hat{\rho}}
  \label{finitexpval}
\end{eqnarray}
with $\hat\rho$ the ensemble density matrix. In equilibrium and when zero
conserved charge is considered the density matrix is given by
\begin{equation}
\hat{\rho} = \exp(-\frac{H}{T}) \label{densmatx}
\end{equation}

In a gauge theory, however, the trace over states in (\ref{finiteteffpot}) must
be defined properly in terms of gauge invariant states. Either the physical
states are selected and only these are used in the trace or alternatively a
projection operator must be introduced in the definition of the
trace\cite{bernard}.

In our approach we select the states as those annihilated by the set of first
class constraints, which are therefore gauge invariant as described in the
previous section.

To one-loop order, we have seen that the Hamiltonian is quadratic in terms of
gauge invariant operators that describe the physical degrees of
freedom. Therefore, to this order the physical partition function is that of a
collection of uncoupled harmonic oscillators for each degree of freedom.

We find the free energy density, which is identified as the finite temperature
effective potential, to be ($\beta=1/T$)
\begin{eqnarray}
{\cal{F}}= V_{eff}(|\chi|;T) & = & V_{eff}(|\chi|;T=0) + \frac{1}{\beta} \int
 \frac{d^3k}{(2\pi)^3} \left\{2 \ln \left[1-e^{-\beta \omega_T(k)}\right] +
 \right. \nonumber \\ & & \left. \ln \left[1-e^{-\beta \omega_H(k)} \right] +
 \ln \left[1-e^{-\beta \omega_p(k) }\right] \right\} \label{finitetpot}
\end{eqnarray}
where $ V_{eff}(|\chi|;T=0)$ is the zero temperature effective potential given
by (\ref{gauginveffpot}), and arises from the zero point energy of the
oscillators.

Just as our gauge invariant approach in terms of gauge invariant operators and
functionals allowed us to obtain the ground state constrained wave functional
eq. (\ref{wavefunctional}), similarly we can obtain the density matrix
elements in the Schroedinger representation $\langle \Phi,\Phi^{\dagger},A_T |
\hat{\rho} | A'_T,{\Phi'}^{\dagger},\Phi'\rangle$.  This representation for
the density matrix is very useful to study non-equilibrium aspects and time
dependent phenomena\cite{frw}, which is the focus of the next section.  We
find the density matrix elements in the Schroedinger representation to be
given by
\begin{equation}
 \langle \Phi,\Phi^{\dagger},A_T | \hat{\rho} | {A'}_T,{\Phi'}^{\dagger},\Phi'
 \rangle = \langle \Phi,\Phi^{\dagger} | \hat{\rho}_{\Phi}
 |{\Phi'}^{\dagger},\Phi' \rangle \otimes \langle A_T | \hat{\rho}_A | {A'}_T
 \rangle,
\end{equation}
where the density matrices are of the harmonic oscillator type, the full
expression is given in the appendix I.

\subsection{Large T expansion:}

The contributions to the free energy from the transverse and Higgs modes are
straightforward to obtain by applying the methods developed by Dolan and
Jackiw\cite{dolan}. However the contribution from the ``plasma'' mode is
non-standard and requires a more detailed analysis which is presented in
appendix II.  We find the leading high temperature behavior to the finite
temperature contribution to the effective potential to be given by
\begin{eqnarray}
V_{eff,T}(|\chi|) & = & -4 \frac{\pi^2 T^4}{90}+
\frac{T^2}{24}\left[3m^2_T+m^2_H+m^2_g\right]
-\frac{T}{12\pi}\left[3(m^2_T)^{\frac{3}{2}}+(m^2_H)^{\frac{3}{2}}+
(m^2_g)^{\frac{3}{2}}\right] + \nonumber \\ & & -\frac{1}{64\pi^2}
\left\{(m^2_g - m^2_T)^2
\ln\left[\frac{m^2_g+m^2_T+2\sqrt{m^2_gm^2_T}}{T^2}\right]+ 2
m^4_T\ln\left(\frac{m^2_T}{T^2}\right)+ \right. \nonumber \\ & & \left.
m^4_H\ln\left(\frac{m^2_H}{T^2}\right) \right\} + \cdots
\label{highTveff}
\end{eqnarray} 

\noindent where the dots stand for terms of ${\cal{O}}(T^0)$ or smaller.
Remarkably, the logarithmic terms cancel similar terms of the zero temperature
part and though this feature is well known in the standard cases (with standard
dispersion relations for the degrees of freedom), it is a new result for the
plasma mode. An important feature of this expression is that the terms linear
in T, which are non-analytic, are {\em complex}.  Whereas the term from the
gauge boson mass is real, the terms originating in the Higgs couplings given by
the contributions from $m^2_H \; , \; m^2_g$ are purely imaginary on the
spinodal regions $|\chi|^2< \mu^2 /3$ for $m^2_H$ and $|\chi|^2< \mu^2 $ for
$m^2_g$. These ``cubic'' terms are usually identified as those responsible for
a first order phase transition and used to compute quantities relevant to the
transition\cite{gleiser1,gleiser2,fodor}.  In particular these terms determine
the supercooling temperature and the latent heat when they are taken as the
leading indicators for a first order transition. In their study of the
electroweak effective potential Anderson and Hall\cite{hall} neglected the
terms involving the Higgs self-coupling keeping only the contributions from the
gauge boson and top quark Yukawa couplings which are gauge invariant and real
to one loop. Arguably such an approximation is justified for very weak Higgs
couplings. Boyd et.al. \cite{brahm} recognized that the terms arising from the
Higgs sector lead to contributions that are imaginary (even after resummation)
precisely as pointed out above.  Thus using these terms to compute the latent
heat, supercooling temperatures and even approximate dynamics is at best a
crude approximation (even when the imaginary parts are ignored) and at worst
disguises other non-equilibrium processes which may be equally important (see
section IV).

\subsection{Comparison with gauge-fixed results:}

In this section we compare our gauge invariant result to the one loop effective
potential obtained in the usual standard path-integral representation for
several gauge-fixing procedures.  The purpose is to {\em contrast} our results
with the suggestion of Fukuda and Kugo\cite{kugo} that there is a large class
of ``good gauges'' for which the effective potential is gauge invariant. These
authors suggested that covariant gauges (including Landau with zero gauge
parameter), $R_\xi$ and other specific gauge fixing schemes are such ``good''
choices.

In order to make a distinction from the gauge invariant formulation, we write
the original complex field $\phi$ (not to be confused with the gauge invariant
field $\Phi$) in the Lagrangian density eq. (\ref{higgslag}) as
\begin{equation}
\phi(\vec x,t) = \frac{1}{\sqrt{2}}\left[\hat{\phi}_R(\vec
x,t)+i\hat{\phi}_I(\vec x,t)\right]+ \varphi
\end{equation}
and $\varphi$ is taken as the (complex) expectation value. To distinguish from
the gauge invariant case we also introduce the masses
\begin{eqnarray}
M^2_g & = & 2\lambda (|\varphi|^2-\mu^2) \label{massgold} \\ M^2_A & = & 2e^2
|\varphi|^2 \label{massphot}\\ M^2_H& = & 2\lambda (3 |\varphi|^2-\mu^2)
\label{masshiggs}
\end{eqnarray}

At this point one would be tempted to identify $\varphi$ with $\chi$, because
the effective tree level masses seem to be the same as in the gauge invariant
case under the replacement $\chi \rightarrow \varphi$. However we make a
distinction between these two expectation values because $\chi$ is a truly
gauge invariant quantity, whereas $\varphi$ is the expectation value of a gauge
transforming field in a fixed gauge.

The one-loop effective potential is given by:

\begin{eqnarray} 
V_1 & = & \sum_{j}\frac{g_j}{2 \beta}\sum_n \int \frac{d^3 k}{(2\pi)^3}
 \ln\left[\left(\frac{2\pi n}{\beta}\right)^2+k^2+M^2_j\right] = V_{10}+V_{1T}
 \nonumber \\ V_{10} & = & \sum_{j}\frac{g_j}{2} \int \frac{d^3 k}{(2\pi)^3}
 \sqrt{k^2+M^2_j}\label{zeroTonelup} \\ V_{1T} & = & \sum_{j}\frac{g_j}{\beta}
 \int \frac{d^3 k}{(2\pi)^3} \ln\left[1-\exp(-\beta \sqrt{k^2+M^2_j}))\right]
 \label{finiteT1lup}
\end{eqnarray}
where the sum are over all particles $j$ with $g_j$ degrees of freedom ($g_j <
0$ for ghosts) and masses $M_j(|\varphi|)$, and we have used a result given in
\cite{dolan}.

The zero temperature contribution is divergent, the $k$ integrals being 
performed with an ultraviolet cutoff $\Lambda$. Discarding a field independent
quartic divergence we find the result
\begin{equation}
 V_{10} = \sum_j \frac{g_j}{4\pi^2} \left\{ \frac{M^4_j}{16}
\ln\left(M^2_j\over\Lambda^2\right) + \frac{\Lambda^2 M^2_j}{4}+
\frac{M^4_j}{32} \right\} \label{1loopunren}
\end{equation}

The finite temperature contribution can be written as
\begin{equation}
 V_{1T} = {T^4\over 2\pi^2}\sum_j g_j I \left( M_j\over T \right)
\label{expfiniteT}
\end{equation}
The high temperature expansion of $I(y)$ is given by\cite{dolan}
\begin{equation}
 I(y) = {-\pi^4\over45} + {\pi^2\over12}y^2 - {\pi\over6}(y^2)^{\frac{3}{2}}-
 {y^4 \over 32} [\ln(y^2) - \frac{3}{2} - {\cal{C}}] + {\cal{O}}(y^6),
 \label{hiTexp}
\end{equation}
where we defined ${\cal C} \equiv 2\ln(4\pi) - 2\gamma = 3.9076$.  The $\ln(
m^2) $ term cancels against the similar term in the zero temperature
contribution $V_{10}$.

\subsubsection{\bf Lorentz gauge:}

Dolan and Jackiw\cite{dolan} calculate the one loop effective potential both at
zero and non-zero temperature in Lorentz gauge, which is ghost free, with gauge
parameter $\alpha$.  The scalar determinant is diagonalized by solving
\begin{equation}
 \ln[k^4 + M_g^2 k^2 + \alpha M_T^2 M_g^2] = \ln[(k^2+M_+^2(\alpha))
(k^2+M_-^2(\alpha))] \label{doljacdet}
\end{equation}
The resulting masses and effective degrees of freedom are
\begin{eqnarray}
M^2_{\pm}(\alpha) = \frac{1}{2}\left[M^2_g \pm \sqrt{M^4_g - 4 \alpha M^2_T
  M^2_g} \right] &\;\;;\;\;& g_{\pm}=1 \label{massplusmin} \\
M^2_g = 2\lambda (|\varphi|^2-\mu^2) &\;\;;\;\;& g_g =1 \label{massgold1} \\
M^2_A = 2e^2 |\varphi|^2 &\;\;;\;\;& g_A=3 \label{massphot1}
\end{eqnarray}

Thus the zero temperature part of the one-loop effective potential in Lorentz
gauge is given by
\begin{eqnarray}
V_{10LG}(|\varphi|;\alpha) & = & \frac{1}{2} \int \frac{d^3k}{(2\pi)^3} \left[
\Omega_H(k)+3 \Omega_A(k)+\Omega_+(k;\alpha)+\Omega_-(k;\alpha)\right]
\label{LGveffT0} \\ \Omega^{\pm}(k;\alpha) & = & \sqrt{k^2+M^2_{\pm}(\alpha)}
\label{omegaplusmin}\\
\Omega_j(k) & = & \sqrt{k^2+M^2_j} \; \; \; ; j=H, \, A. \label{bigomega}
\end{eqnarray}

The gauge dependence (dependence on the gauge parameter $\alpha$) of the above
result is explicit. Even for Landau gauge, that is $\alpha=0$, expression
(\ref{LGveffT0}) describes {\em{five}} degrees of freedom, rather than the
four physical degrees of freedom described by the gauge invariant result
(\ref{gauginveffpot}).  With the same renormalization prescription leading to
the gauge invariant result eq. \ref{veffren} we find in Landau gauge the
result
\begin{eqnarray}
V(\varphi,\alpha=0) & = & \lambda (|\varphi|^2-\mu^2)+ \frac{1}{4\pi^2}\left\{
 \frac{1}{32}\left[3M^4_A+M^4_H+M^4_g \right] + \right. \nonumber \\ & &
 \left. \frac{1}{16} \left[ 3M^4_A \ln(\frac{m^2_T}{\kappa^2})+ M^4_H
 \ln(\frac{M^2_H}{\kappa^2})+ M^4_g \ln(\frac{M^2_g}{\kappa^2})\right] \right\}
 \label{alfa0}
\end{eqnarray}
which is obviously very different from the gauge invariant result given by
eq. \ref{veffren} if $\varphi$ is identified with the gauge invariant order
parameter $\chi$.

The effective potential (\ref{LGveffT0}) becomes independent of the gauge
parameter $\alpha$ for the values $\varphi = 0 \; ; |\varphi|^2= \mu^2$. These
are the value of the extrema of the {\em tree level} potential.  The gauge
dependence appears at 1-loop order and is therefore formally of
${\cal{O}}(\hbar)$ since the extrema of the effective potential will acquire
${\cal{O}}(\hbar)$ corrections. We identify the values of $\varphi$ at which
the gauge dependence cancels out as the extrema of the effective action {\em
to this order}.  Up to an irrelevant constant the gauge invariant effective
potential (\ref{gauginveffpot}) and the one-loop effective potential in
general covariant gauge (\ref{LGveffT0}) are the same for $|\chi|^2 = \mu^2$,
i.e. at the extrema of the effective action. This equality is a consequence of
the known result that the extrema of the effective action are gauge
independent. At zero temperature gauge independence at the extrema is also a
consequence of the Nielsen identities\cite{nielsen}.  Therefore we find that
the suggestion of reference\cite{kugo} is {\em only} valid at the extrema of
the effective action.

At finite temperature we find that the ${\cal{O}}(T^2)$ is gauge parameter
independent and given by
\begin{equation}
V_{1T}^{(2)} = \frac{T^2}{24} \left[3M^2_A+M^2_H+M^2_g \right]
\label{T2contLG}. \\
\end{equation}
This term coincides with that from the gauge invariant effective potential with
the identification $|\varphi| \equiv |\chi|$.  This contribution determines (to
this order) the critical temperature which is therefore a gauge independent
quantity, whereas the ${\cal{O}}(T)$ and ${\cal{O}}(\ln(T))$ contributions are
gauge parameter dependent.

The ${\cal{O}}(T)$ contribution is given by
\begin{equation}
V_{1T}^{(1)} = -\frac{T}{12\pi}\left[3(M^2_A)^{\frac{3}{2}}+
(M_H^2)^{\frac{3}{2}}+ (M^2_{+}(\alpha))^{\frac{3}{2}}+
(M^2_{-}(\alpha))^{\frac{3}{2}}\right] \label{T1contLG}
\end{equation}
For $\alpha=0$, $M_+^2=M^2_g \; ; \; M_-^2=0$ and this ${\cal{O}}(T)$ term
coincides with the ${\cal{O}}(T)$ contribution from the gauge invariant
effective potential given by eq. (\ref{highTveff}) if $|\varphi|$ is identified
with the gauge invariant order parameter $|\chi|$.

This contribution is of particular importance because it is usually taken as a
signal for a first order transition and determines its strength, and sometimes
used in phenomenological equations to describe the
dynamics\cite{gleiser1,gleiser2}.  In the case of first order phase
transitions, this term is sometimes used to compute the latent heat and the
supercooling temperature\cite{fodor}. Clearly, quantities calculated
solely from this term would be physically meaningless because of the gauge
dependence.  Furthermore, this contribution is complex for $|\varphi|^2 <
\mu^2$; only the contribution from the gauge boson is real and gauge invariant.
Therefore our conclusion is that this term is provides the correct gauge
invariant ${\cal{O}}(T)$ contribution {\em only} in Landau ($\alpha=0$) gauge
and with the identification $|\varphi| \equiv |\chi|$. However, this
equivalence only holds to leading order in the high temperature expansion and
{\em is not} a general feature to all orders, as displayed explicitly by higher
order finite temperature corrections and also by the zero temperature part.

\subsubsection{\bf $R_\xi$ and $\bar{R}_\xi$ gauges:} 

In $R_\xi$ gauge the following gauge fixing and ghost terms are added to the
Lagrangian density
\begin{equation}
 {\cal L}_{GF} = -\frac{1}{2\xi} (\partial_\mu A^\mu + \xi e \varphi
    \hat{\phi}_I)^2, \qquad {\cal L}_{FPG} = c^\dagger \left[-\partial^2 - \xi
    e^2 \varphi (\varphi+\hat{\phi}_R) \right] c
\end{equation} 

Kastening \cite{bkast} describes a useful variant called $\bar{R}_\xi$ gauge,
in which $\varphi$ and $\hat{\phi}_R$ are treated more symmetrically.  Its
gauge fixing and ghost terms are:
\begin{equation}
 {\cal L}_{GF} = -\frac{1}{2\xi} \left[\partial_\mu A^\mu + \xi e
 (\varphi+\hat{\phi}_R) \hat{\phi}_I\right]^2, \qquad {\cal L}_{FPG} =
 c^\dagger \left[-\partial^2 - \xi e^2 [(\varphi+\hat{\phi}_R)^2 -
 \hat{\phi}^2_I] \right] c
\end{equation}
The ghost term is derived as usual by looking at the response of the gauge
fixing functional under a gauge transformation. The corresponding gauge fixing
functional for the $\bar{R}_\xi$ gauge fixing procedure is $f[A^\mu] =
\partial_\mu A^\mu + \xi e {i\over2} (\phi^\dagger \phi^\dagger - \phi \phi)$.
A piece from the photon determinant cancels off half the ghost contribution, so
we will treat the ghosts as having $g_c=-1$.  In both $R_\xi$ and $\bar{R}_\xi$
gauges we find the following masses and degrees of freedom:
\begin{eqnarray}
M^2_A &\;\;;\;\;& g_A=3 \\ M^2_c = \xi M^2_A &\;\;;\;\;& g_c=-1 \\ M^2_H
  &\;\;;\;\;& g_H = 1 \\ M^2_\xi = M^2_g + \xi M^2_A &\;\;;\;\;& g_\xi=1
\end{eqnarray} 

The degree of freedom with $M_\xi$ is a (gauge-dependent) linear combination
of the Goldstone mode and the vector boson. At the tree level minimum
$|\varphi|^2=\mu^2$ when $M^2_g=0$ this mode cancels the remaining ghost
contribution, leaving a $\xi-$ independent result, calculable just from the
photon and the Higgs terms. At the tree level maximum $\varphi=0$ the gauge
dependence cancels out completely and the ghost decouples.  Away from the
minimum, however, the 1-loop effective potential is $\xi-$ dependent.

At finite temperature we find that the ${\cal{O}}(T^2)$ term is again gauge
parameter independent and given by the same expression as in equation
(\ref{T2contLG}).

However the next, ${\cal{O}}(T)$, term is explicitly gauge parameter dependent
and given by
\begin{equation}
 {-T\over12\pi} [3(M_A^2)^{3/2}) - (\xi M_A^2)^{3/2} + (M_H^2)^{3/2} + (M_g^2 +
 \xi M_A^2)^{3/2}] \label{ordTxigau}
\end{equation}
This term depends on $\xi$
(unless $e=0$ or $|\varphi|^2=\mu^2$) and is complex for $|\varphi| < \mu^2$.
In Landau gauge ($\xi=0$), it is given by
\begin{equation}
 {-T\over12\pi} [3(M_A^2)^{3/2} + (M_H^2)^{3/2} + (M_g^2)^{3/2}] \qquad
\hbox{(Landau)} \label{landaugauge}
\end{equation}
and coincides with the
${\cal{O}}(T)$ contribution to the high T expansion of the gauge invariant
effective potential (\ref{highTveff}) if $\varphi$ is identified with the gauge
invariant order parameter $\chi$.

Since to this order only the ${\cal{O}}(T^2)$ term enters in the estimate of
the critical temperature, we see that $T_c$ is a gauge invariant quantity (to
this order). But comparing the ${\cal{O}}(T)$ term with that in Lorentz gauges
should convince the reader that this term is gauge dependent and complex in
general and one must be very careful in attaching any physical meaning to it,
such as a criterion for a first order phase transition, its strength and the
ensuing latent heat and supercooling temperature. Only in Landau gauge do we
find that this term is the same as in the gauge invariant formulation, although
even in this gauge, the higher order finite temperature corrections and the
zero temperature part are gauge parameter dependent.

\subsubsection{\bf Unitary Gauge}

In unitary gauge, even the leading term of $V_{1T}$ (calculated to 1 loop) is
incorrect: \cite{dolan}
\begin{equation}
 {T^2\over24} [3(M_A^2) + M_H^2] \label{unitaryga}
\end{equation}
In this
gauge, higher-loop corrections affect the leading term and must be included in
the calculation to obtain the correct answer (and gauge independent) for the
${\cal{O}}(T^2)$ term.  Arnold, Braaten and Vokos \cite{abv} showed that a
2-loop calculation restores the correct leading term (\ref{T2contLG}).

\subsubsection{\bf Higher-Loop Resummation}

We have seen that the 1-loop effective potential, calculated in various
gauge-fixing schemes, is explicitly gauge parameter-dependent.  One might argue
that a resummation of higher-loop diagrams would eliminate this dependence.
Here we will show that this is not the case\cite{corr}.

Although resummation would naturally be invoked to restore gauge invariance in
a particular calculation, clearly resummation does nothing to restore gauge
parameter-independence at $T=0$, since higher-loop diagrams are all
higher-order in $\lambda$ or $e^2$. Thus it is clear that the gauge dependence
will remain in any calculation that includes the zero temperature contribution.

At finite (but large) $T$ it is a little less obvious, since there is an
additional expansion parameter: $M/T$ (with $M$ any of the masses).  In fact,
we have already mentioned that resummation in unitary gauge {\it does\/}
recover the correct, gauge-independent, $T^2$ term.  We now show explicitly in
$R_\xi$ and $\bar{R}_\xi$ gauges that the $T$ term remains $\xi$-dependent
even after resummation of hard thermal loops\cite{corr}.

{\bf Expansion Parameters:}

In this section we take $e^2 \sim \lambda$, and $M$ can represent $M_H$ or
$M_A$.  Terms in $V_T$ are described as ${\cal O}(\alpha^a \beta^b
\gamma^c)$ with respect to the leading $T^2 M^2$ term, where \cite{corr,eqz1}
\begin{equation}
 \alpha \equiv \lambda T^2/m^2, \qquad \beta \equiv \lambda T/m, \qquad
  \gamma \equiv \varphi^2/T^2
\end{equation}
We will take $\alpha\approx 1$ and $\gamma\approx 1$ but $\beta<1$, so that
$\{e^2,\lambda\} = {\cal O}(\beta^2)$ and $M/T = {\cal O}(\beta)$.  Our
1-loop high-$T$ expansion is then seen to be an expansion in $\beta$ (not to
be confused with inverse temperature here).

We want to resum to ${\cal O}(\beta)$ (given by daisy diagrams with hard
thermal loops).  The simplest approach is the tadpole method (for a more
comprehensive discussion, see ref.\cite{corr}), replacing the $M_j$'s by
thermal masses, which need only be calculated to ${\cal O}(\beta^0)$.  We then
integrate with respect to $\varphi$, adding constants to taste, to give $V_T$.
The leading ($T^2$) term is unaffected, and remains gauge-independent.

{\bf Thermal Masses:}

The ${\cal O}(\beta^0)$ thermal masses are unambiguous and gauge-independent.
For the scalars we can use the relations $\tilde M_H^2 = M_H^2 + V''_{1T}
(\varphi)$ and $\tilde M_\xi^2 = M_\xi^2 + V'_{1T}(\varphi)/ \varphi$ (the
former is true to all orders in $\bar{R}_\xi$ gauge).  We get
\begin{equation}
 \tilde M_H^2 = M_H^2 + {T^2\over24} [6 e^2 + 8 \lambda], \qquad \tilde M_\xi^2
  =M_g^2 + \xi M_A^2 + {T^2\over24} [6 e^2 + 8 \lambda]
\end{equation}
Only the longitudinal part of the photon gets a thermal (electric screening,
plasma) mass (given by $\Pi_0^0$) to leading order:
\begin{equation}
 \tilde M_L^2 = M_A^2 + {e^2 T^2 \over 3}, \qquad \tilde M_T^2 = M_A^2
\end{equation}

With these hard-thermal loop resummed masses, the ${\cal O}(\beta)$ term in
$R_\xi$ or $\bar{R}_\xi$ gauge becomes
\begin{eqnarray}
  {-T\over12\pi} &\!\!& \left[ 2 M_A^3 + (1-\xi^{3/2}) \left( M_A^2 + {e^2
    T^2\over 3} \right)^{3/2} + \left( M_H^2 + {T^2 \over 24} [6 e^2 + 8
    \lambda] \right)^{3/2}\right. \nonumber\\
  &+& \left. \left( M_g^2 + \xi M_A^2 + {T^2\over 24} [6 e^2 + 8 \lambda]
    \right)^{3/2} \; \right] \label{rxirs}
\end{eqnarray}
This term is still $\xi$-dependent.  We see that resummation (to ${\cal
O}(\beta)$) does not render the effective potential gauge
parameter-independent.

\section{\bf Non-equilibrium aspects}

The main motivation for studying an effective potential is to address the issue
of symmetry breaking and phase transitions accounting for quantum and thermal
corrections. By its very definition, the effective potential is an
{\em{equilibrium}} quantity because the expectation value of the scalar field,
that serves as the order parameter, is space-time independent. At zero
temperature the quantum wave-functional is taken to be a stationary state, at
finite temperature, the calculation is performed with the equilibrium partition
function. Thus it is clear that the effective potential is only suitable to
describe the equilibrium aspects associated with the phase transition. The only
equilibrium states correspond to the extrema of the effective action which for
homogeneous configurations coincide with the extrema of the effective
potential. Thus quantities like the critical temperature defined as the value
at which the minima of the effective potential become a maximum, as well as
expectation values of the scalar field that extremize the effective potential
are meaningful quantities that are useful to determine whether there is a
symmetry breaking phase transition.

For values of the order parameter away from the extrema, the effective
potential is simply {\em{not}} a reliable tool to describe the situation and by
its very definition it is not meant to be. In using the effective potential to
address dynamical issues, like ``the rolling down'' of the order parameter
towards the equilibrium configuration the hope is that the time evolution is
rather slow and the use of the ``instantaneous'' effective potential is
justified as some sort of adiabatic approximation. This expectation, however,
conceals the physics of the dynamical processes during the phase
transition\cite{danvega}.

In second order phase transitions, if the order parameter is initially very
small and the system is evolving from an initial disordered (high temperature)
phase with short range correlations towards a final equilibrium broken symmetry
state with long range correlations, the dynamical process is that of phase
separation and growth of correlated domains. This dynamics is not captured
solely by the time evolution of the expectation value of the scalar field, but
the growth of fluctuations will be manifest in the time dependence of the
correlation functions of this field. During the early stages of a second order
phase transition, when the order parameter is very small (near the maximum of
the effective potential) small amplitude, long wavelength fluctuations become
unstable and grow\cite{domain}. The two-point correlation functions of the
scalar field reflect this instability and grow exponentially, as
the order parameter ``rolls down'' the potential hill.

If the phase transition is of first order, then there are free energy barriers
that the system has to overcome to reach the equilibrium state and the phase
transition is driven by nucleation, in which large amplitude configurations
become unstable and grow. However, if the first order transition is very weak
(that is to say that there is a substantial amount of phase mixing) nucleation
and phase separation occur on similar spatial and time scales and the
long-wavelength instabilities will still be important.

The information on thermodynamic instabilities that lead to phase separation is
{\em contained} in the effective potential both at zero and finite temperature
in the form of an {\em imaginary part}.

Weinberg and Wu\cite{weinberg} have shown in a beautiful paper, that the
imaginary part of the effective potential in scalar theories determines the
growth rate of the scalar field two-point correlation function for Gaussian
states centered at zero expectation value (in field space). Boyanovsky and de
Vega\cite{danvega} have studied how this growth of correlations affects the
time evolution of the order parameter and concluded that the use of the
effective potential to describe the time evolution of the expectation value of
the scalar field is not only unwarranted but completely misleading and
unreliable whenever the initial value of the expectation value is in the
``spinodal region'' (the region in the tree-level potential where the second
derivative is negative).

\subsection{\bf The Imaginary Part}

We notice that in the ``classical spinodal'' region, where $m^2_H <0$ or 
$0< |\chi| <\mu / \sqrt{3}$, the Higgs mode has a band of unstable wave-vectors
with imaginary frequencies for $k^2< \mu^2-3|\chi|^2$. The plasma mode has a
band of unstable wave vectors with imaginary frequencies for $k^2<
\mu^2-|\chi|^2$ in the spinodal region $0<|\chi|^2< \mu^2$. This new spinodal
line ranges from the maximum to the minima of the tree level potential.

When the gauge coupling is switched off, in the absence of long-range forces,
this is recognized as the ``spinodal'' region for the would be Goldstone
modes\cite{frw}. However in the presence of long range forces, the
instabilities are much more severe for long-wavelength fluctuations, as can be
seen from the infrared behavior of the plasma frequency eq. (\ref{plasmafreq}).

As a consequence of these unstable modes, both the zero and finite temperature
effective potential acquire an imaginary part for {\em all values} of $|\chi|$
between the maximum and the minima of the tree level potential. Therefore we
see that unlike the case of a scalar order parameter in which there is a
thermodynamically stable region in the phase diagram (between the classical
spinodal and the coexistence line), in this case the unstable region covers the
whole of the phase diagram below the coexistence curve.

Thus we here obtain one of the important conclusions of this work: the gauge
invariant effective potential to one-loop order is {\em {complex}} from the
maximum to the minimum of the tree-level potential. There are no homogeneous
configurations corresponding to quasi-equilibrium states away from the minimum.
Explicitly, the imaginary part of the gauge invariant effective potential at
zero temperature is given by:
\begin{eqnarray}
Im(V_{eff}) = && \pm \frac{1}{4\pi^2} \left\{ \int_0^{k_1(|\chi|)}k^2 dk
\sqrt{k^2_1(|\chi|) - k^2}\quad \Theta(\mu^2 - 3|\chi|^2) \right.  \nonumber \\
&+& \left. \int_0^{k_2(|\chi|)}k dk \sqrt{k^2_2(|\chi|) -
k^2}\sqrt{2e^2|\chi|^2+ k^2} \quad \Theta(\mu^2 - |\chi|^2) \right\} \nonumber
\\ && {k^2_1(|\chi|)}= 2\lambda (\mu^2 - 3|\chi|^2) ; \quad\quad\quad
{k^2_2(|\chi|)}= 2\lambda (\mu^2 - |\chi|^2) \label{impartvef}
\end{eqnarray}
where the $\pm$ is determined by the direction of the analytic continuation in
the frequencies.

The presence of an imaginary part of the effective potential has been sometimes
justified as a failure of the loop expansion and ``corrected'' by the Maxwell
construction which yields a convex free energy.  However, just as the Van der
Walls loop in the equation of state for liquid-gas systems is a signal of
thermodynamic instabilities, the imaginary part of the effective potential
signals the presence of similar instabilities in the quantum system.

The flat region of the Maxwell constructed free energy indicates that the
system will be found in a coexistence of phases but offers no information on
the non-equilibrium processes leading to phase separation. In this respect the
effective potential with its imaginary part at least provides some restricted
information on dynamical processes. As discussed in
references\cite{weinberg,danvega} this imaginary part determines the decay rate
of unstable Gaussian states.

In the one-loop approximation, the Hamiltonian for the modes with wavevectors
in the unstable bands correspond to {\it {inverted harmonic oscillators}} for
which the analysis of references\cite{weinberg,danvega,frw,guthpi} can be
applied.  The effective potential (\ref{gauginveffpot}) is complex because the
modes corresponding to inverted harmonic oscillators were treated as ordinary
harmonic oscillators, i.e. by {\em analytically continuing} the zero point
energy for these oscillators. The Gaussian wavefunctionals and density matrix
have complex kernels for the fields $\eta_{1,2}$ reflecting this analytic
continuation.

In terms of the shifted fields $\eta_1\; , \eta_2$ with zero expectation value,
the Gaussian wave-functional (\ref{wavefunctional}) and the density matrix (see
appendix I) are ``centered'' at the origin in field space, and for the
wavevectors in the unstable bands, they are not stationary states of the
Hamiltonian.

In this one-loop approximation we can use the results of previous
investigations of similar situations\cite{weinberg,danvega,frw,guthpi} to study
the early time evolution of initially prepared non-equilibrium states.

\subsection{\bf Early time evolution of non-equilibrium states}

{}From the previous discussion it is clear that the imaginary part of the
effective potential conceals a dynamical non-equilibrium situation associated
with the instabilities towards phase separation. This is a dynamical situation
that {\em must} be treated as a time dependent problem.  The quantum states
that lead to the effective potential (and free energy) are {\em not} stationary
states.

Since only the Higgs and plasma modes have unstable bands we will focus only on
these.  Consider the situation in which an initial Gaussian density matrix has
been prepared such that the gauge invariant scalar field $\Phi$ has initial
expectation value in this ensemble given by $\chi$, and that the kernels have
positive frequencies.  That is, the density matrix in the Schroedinger
representation is given by the expression in appendix I but with the following
real kernels:
\begin{eqnarray}
&&K^{(1)}_H(k,t=0) = -\frac{1}{2}W_H(k)\coth\left[\frac{W_H(k)}{T}\right] \; \;
; \; \; K^{(2)}_H(k,t=0)= \frac{W_H(k)}{\sinh\left[\frac{W_H(k)}{T}\right]}
\label{cova1} \\ &&K^{(1)}_p(k,t=0) =
-\frac{1}{2}W_P(k)\coth\left[\frac{W_P(k)}{T}\right] \; \; ; \; \;
K^{(2)}_H(k,t=0)= \frac{W_P(k)}{\sinh\left[\frac{W_P(k)}{T}\right]}
\label{cova2} \\ && W^2_H(k) >0 \; \; , \; \; W^2_P(k) > 0 \label{posfreq}
\end{eqnarray}
The frequencies $W_H(k)\; , \; W_P(k)$ determine the initial conditions on the
state as discussed in refs.\cite{weinberg,danvega,frw,guthpi}.  We can now use
the results of references\cite{weinberg,danvega,frw,guthpi} to determine the
time evolution of this initial state.  The early time evolution of the
expectation value $\chi$ will be determined by the classical equations of
motion with small quantum and thermal corrections. However at longer times the
growth of the unstable modes makes the one-loop approximation to the
dynamics unreliable  as in the problem of domain growth in scalar
theories\cite{frw,domain}.  Clearly, the most important dynamics is described
by the evolution of the unstable modes and determined by the time evolution of
the kernels.  We now use the method discussed in reference\cite{frw} to obtain
the following time evolution of the kernels:

\begin{eqnarray}
&& K^{(1)}_H(k,t)= \frac{ K^{(1)}_H(k,t=0) }{|U_k(t)|^2} -i \frac{d
 \ln(|U_k(t)|^2}{dt} \label{K1Hoft} \\ && K^{(2)}_H(k,t) =\frac{
 K^{(2)}_H(k,t=0) }{|U_k(t)|^2}\label{K2Hoft} \\ && K^{(1)}_p(k,t)= \frac{
 K^{(1)}_p(k,t=0) }{|V_k(t)|^2} -i \frac{d \ln(|V_k(t)|^2}{dt} \label{K1poft}
 \\ && K^{(2)}_p(k,t) =\frac{ K^{(2)}_p(k,t=0) }{|V_k(t)|^2}\label{K2poft}
\end{eqnarray}
where the mode functions $U_k(t) = U_{1k}(t)+iU_{2k}(t) \; , \;
V_{1k}(t)+iV_{2k}(t)$ obey the following evolution equations\cite{frw}
\begin{eqnarray}
&& \left[\frac{d^2}{dt^2}+ k^2+2\lambda(3|\chi_{cl}|^2(t)-\mu^2)\right]U_k(t) =
0 \nonumber \\ && U_{1k}(0)=1 \; , \; U_{2k}(0)=0 \; , \; \dot{U}_{1k}(0)=0 \;
, \; \dot{U}_{2k}(0)= W_H(k) \label{umodeq} \\ && \left[\frac{d^2}{dt^2}+
\left[k^2+2\lambda(|\chi_{cl}|^2(t)-\mu^2)\right] \left[
k^2+2e^2|\chi_{cl}|^2(t)\right]/k^2 \right]V_k(t) = 0 \nonumber \\ &&
V_{1k}(0)=1 \; , \; V_{2k}(0)=0 \; , \; \dot{V}_{1k}(0)=0 \; , \;
\dot{V}_{2k}(0)= W_P(k) \label{vmodeq}
\end{eqnarray}
where $\chi_{cl}$ is the classical evolution of the ``zero mode''\cite{frw}.
Obviously the mode functions $U_k(t) \; , \; V_k(t)$ grow almost exponentially
for wave-vectors in the respective unstable bands when $|\chi_{cl}|(t)$ is in
the corresponding spinodal regions.

Consider the equal-time two-point correlation functions of the {\em gauge
invariant} operator $\eta,\ \eta^{\dagger}$:
\begin{equation}
\langle \eta(\vec{x},t) \eta^{\dagger}(\vec{y},t) \rangle = \int
\frac{d^3k}{(2\pi)^3}e^{i\vec{k}\cdot (\vec{x}-\vec{y})}
\left\{\frac{|U_k(t)|^2}{2W_H(k)} \coth\left[\frac{W_H(k)}{T}\right]+ \right.
\left. \frac{|V_k(t)|^2}{2W_P(k)} \coth\left[\frac{W_P(k)}{T}\right] \right\}
\label{corrfunct}
\end{equation}
When $\chi_{cl}(t)$ remains in the spinodal, it is clear that the {\em growth
rate} of $\langle \eta(\vec{0},t) \eta^{\dagger}(\vec{0},t) \rangle$ is related
to the imaginary part of the effective potential given by eq. (\ref{impartvef})
at zero temperature.

At this point we recognize an important payoff of the gauge invariant
description. In order to compute gauge invariant correlation functions from the
gauge variant operator $\phi(\vec x, t)$ we would have to append a line
integral of the (time dependent) gauge field, with the ensuing path ambiguities
and complications. The formulation in terms of gauge invariant order
parameters from the start overcomes these difficulties and allows to extract
physically meaningful correlation functions that provide dynamical information
on non-equilibrium processes.

Just as in the case of a scalar field theory undergoing spinodal
decomposition\cite{domain} this equal-time correlation function will grow
exponentially at early times because of the unstable modes. This growth of is
the hallmark of the dynamics of phase separation\cite{weinberg,danvega,domain}.

If the phase transition is weakly first order in the sense that a considerable
amount of phase separation and mixing occurs during nucleation, then the growth
of correlations described by the dynamics of the unstable modes will be also a 
dominant process.

As noted before, unlike the case of a scalar field transforming under a
discrete symmetry, in this case the spinodal region reaches all the way to the
minimum of the potential (at least in the one-loop approximation) and the
long-wavelength instabilities are enhanced by the long-range forces. This
feature provides interesting possibilities if the phase transition is strongly
first order and nucleation is the dominant mechanism. Typically after
nucleation of a critical bubble, the order parameter inside the bubble is very
close to its equilibrium value. If this value is smaller in magnitude to that
of the equilibrium configuration, spinodal instabilities may be still present
{\em inside} the bubble.

If the critical droplet has rather small radius it will have to grow to a size
$\approx 1/(|\chi|-\mu)$, with $|\chi|$ the value of the order parameter inside
the bubble, before any instability develops inside the bubble. Therefore the
fact that the spinodal line reaches to the minimum may lead to the tantalizing
possibility that for large critical bubbles there may still be non-equilibrium
processes {\em inside} the bubble if the order parameter inside is smaller in
magnitude than the equilibrium value.

Clearly all these possibilities will have to be studied in deeper detail and we
expect a strong dependence on the values of the gauge and Yukawa couplings (in
the case of fermions). Furthermore our analysis of the time evolution of the
spinodal instabilities only holds at very early times after the preparation of
the initial state. At longer times a non-perturbative gauge invariant scheme
will have to be used to determine the dynamics. We hope to be able to implement
the variational schemes of reference\cite{camelia} to the gauge invariant
formulation.

\section{Conclusions}

In this article we have presented a formulation of the effective potential in
terms of a gauge invariant order parameter for the case of the Abelian Higgs
model.  The gauge invariant states of the theory are those annihilated by the
first class constraints and gauge invariant operators are those that commute
with these constraints. We recognized an order parameter that is invariant
under the local gauge transformations but transforms as a charged operator
under global phase rotations, its expectation value in the lowest energy state
therefore signals the breakdown of the global $U(1)$ symmetry.

We evaluated the one-loop contribution to this gauge invariant effective
potential both at zero and non-zero temperature and obtained its high
temperature expansion. We found that the effective potential is complex both at
zero and non-zero temperature and that the spinodal (thermodynamically
unstable) region extends from the maximum to the minimum of the tree level
potential (in this approximation).  The gauge invariant effective potential was
compared to the effective potential obtained in several covariant gauges, we
found that the dependence on gauge parameter cancels only at the extrema, but
for all other values of the order parameter the gauge fixed effective
potentials are gauge parameter dependent at zero and non-zero temperature. In
particular in a high temperature expansion only the ${\cal{O}}(T^2)$
contribution is gauge invariant, whereas the ${\cal{O}}(T)\; , \;
{\cal{O}}(\ln(T)) \cdots $ depend on the gauge parameter and are complex. In
general these cannot be taken as trustworthy quantities to extract information
on the strength of the transition and its features, such as latent heat and
supercooling temperature, although the contribution of the gauge boson masses
(we did not study fermions) are gauge independent and real to this order.

The imaginary part of the one-loop effective potential determines the spinodal
lines, which identify the region of long-wavelength instabilities.  Unlike the
case of a scalar field with discrete symmetries, the spinodal line encompasses
all values of the homogeneous order parameter ranging between the maximum and
the minimum of the tree level potential in the one-loop approximation.

Non-equilibrium dynamical aspects were then studied in terms of gauge invariant
correlation functions. It is in these dynamical situations out of equilibrium
that the gauge invariant formulation has the greatest impact, since in general,
correlation functions involve non-local line integrals of the gauge field.  We
obtained the early time behavior of the gauge invariant correlation functions
for initial conditions in which the order parameter is in the spinodal region,
these correlation functions grow exponentially at early times as a consequence
of the instabilities.

We conjectured that these spinodal instabilities may play an important role in
first order phase transitions, in that they may be responsible for
non-equilibrium dynamics {\em inside} the nucleating bubbles. Clearly this
possibility will have to be studied further.

There are several important avenues to pursue: higher order calculations and
the implementation of variational or Hartree-like approximations in the gauge
invariant formulation to address the long-time behavior and to obtain a more
solid understanding of non-equilibrium processes. Furthermore, it is important
to generalize the resummation program to the gauge invariant effective
potential, and to generalize these new methods and results to non-Abelian gauge
theories in particular the electroweak theory. Study on these issues is
underway.

\section{Appendix I}

As shown in section II, in the one-loop approximation the Hamiltonian becomes a
sum of independent harmonic oscillators for the gauge invariant normal modes
and in this case the explicit form of the density matrix in the Schroedinger
representation is available (see\cite{frw} and references therein).  This form
of the density matrix is particularly convenient to study real-time dynamics,
since the Liouville equation becomes a functional differential equation, which
in this case can be solved explicitly for the time dependence\cite{frw}.

The {\em equilibrium} density matrix that describes the system at temperature
$T= 1/\beta$ is given in the Schroedinger representation by:
\begin{eqnarray}
\lefteqn{ \langle \Phi,\Phi^{\dagger},A_T | \hat{\rho} |
  {A'}_T,{\Phi'}^{\dagger},\Phi' \rangle = N[T] \times} \nonumber\\
\lefteqn{\exp\left\{\int\!\! d^3x \int\!\! d^3y \left[\vec{A}_T(\vec{x}) \cdot
  \vec{A}_T(\vec{y}) +\vec{A'}_T(\vec{x}) \cdot \vec{A'}_T(\vec{y}) \right]
  K^{(1)}_T(\vec{x}-\vec{y}) +\vec{A'}_T(\vec{x}) \cdot \vec{A}_T(\vec{y})
  K^{(2)}_T(\vec{x}-\vec{y}) \right\}} \nonumber\\
&\times& \exp\left \{ \int\!\! d^3x \int\!\! d^3y \left[ \eta_1(\vec{x})
  \eta_1(\vec{y}) + \eta'_1(\vec{x}) \eta'_1(\vec{y})\right]
  K^{(1)}_H(\vec{x}-\vec{y})+ \eta'_1(\vec{x}) \eta_1(\vec{y})
  K^{(2)}_H(\vec{x}-\vec{y})\right\} \nonumber\\
&\times& \exp\left \{ \int\!\! d^3x \int\!\! d^3y \left[ \eta_2 (\vec{x})
  \eta_2(\vec{y}) + \eta'_2 (\vec{x}) \eta'_2(\vec{y}) \right]
  K^{(1)}_p(\vec{x}-\vec{y}) + \eta'_2(\vec{x}) \eta_2(\vec{y})
  K^{(2)}_p(\vec{x}-\vec{y}) \right\}
\end{eqnarray}
\begin{eqnarray}
N [T] &=& N[0] \Pi_{k}\left[\tanh^2\left(\frac{\omega_T(k)}{T}\right)
  \tanh\left(\frac{\omega_H(k)}{T}\right)
  \tanh\left(\frac{\omega_p(k)}{T}\right) \right]^{\frac{1}{2}} \\
K^{(1)}_T(\vec{x}-\vec{y}) &=& -\frac{1}{2}\int \frac{d^3k}{(2\pi)^3}
  \omega_T(k) \coth[\frac{\omega_T(k)}{T}] e^{i \vec{k} \cdot
  (\vec{x}-\vec{y})} \\
K^{(2)}_T(\vec{x}-\vec{y}) &=& \int \frac{d^3k}{(2\pi)^3} \frac{\omega_T(k)}
  {\sinh[\frac{\omega_T(k)}{T}] } e^{i \vec{k} \cdot (\vec{x}-\vec{y})} \\
K^{(1)}_H(\vec{x}-\vec{y}) &=&
  -\frac{1}{2} \int \frac{d^3k}{(2\pi)^3} \omega_H(k)
  \coth[\frac{\omega_H(k)}{T}] e^{i \vec{k} \cdot (\vec{x}-\vec{y})} \\
K^{(2)}_H (\vec{x}-\vec{y}) &=& \int \frac{d^3k}{(2\pi)^3}
  \frac{\omega_H(k)} {\sinh[\frac{\omega_H(k)}{T}] } e^{i \vec{k} \cdot
  (\vec{x}-\vec{y})} \\
K^{(1)}_p (\vec{x}-\vec{y}) &=&
  -\frac{1}{2} \int \frac{d^3k}{(2\pi)^3} \frac{k^2 \omega_p(k)
  }{\omega^2_T(k)} \coth\left[ \frac{k^2 \omega_p(k) }{T \omega^2_T(k)}\right]
  e^{i \vec{k} \cdot (\vec{x}-\vec{y})} \\
K^{(2)}_p (\vec{x}-\vec{y}) &=& \int \frac{d^3k}{(2\pi)^3} \frac{\omega_p(k)
  k^2} {\omega^2_T(k)\sinh\left[\frac{k^2 \omega_p(k) } {T
  \omega^2_T(k)}\right]} e^{i \vec{k} \cdot (\vec{x}-\vec{y})}
\end{eqnarray}

The trace of this density matrix gives the gauge invariant one-loop partition
function, whose logarithm (divided by $-\beta$) gives the one-loop effective
potential at finite temperature.

\section{Appendix II}

In this appendix we provide the essential ingredients to obtain the high
temperature expansion of the contribution of the plasma mode to the finite
temperature effective potential.

The first two terms in the free energy density given by eq. (\ref{finitetpot})
(the transverse and Higgs modes) are of the usual form with dispersion
relations of the form $\omega(k) = \sqrt{k^2+m^2}$ and their asymptotic high
temperature expansion can be obtained by following the steps described
in\cite{dolan}.  The third term requires special attention because of the
unusual dispersion relation of the plasma mode. This term can be written as
\begin{eqnarray} 
 \int \frac{d^3k}{(2\pi)^3} \ln \left[1-e^{-\beta \omega_p(k) }\right]& = &
 \frac{T^4}{2\pi^2} I(y^2_1,y^2_2) \nonumber \\ I(y^2_1,y^2_2) & = &
 \int^{\infty}_0 x^2 dx
 \ln\left[1-e^{-\frac{1}{x}\sqrt{(x^2+y^2_1)(x^2+y^2_2)}}\right] \nonumber \\
 y^2_1 & = & \frac{m^2_g}{T^2} \; \; \; y^2_1 = \frac{m^2_T}{T^2} \label{Iterm}
\end{eqnarray} 

The function $ I(y^2_1,y^2_2)$ can be written as
\begin{equation}
 I(y^2_1,y^2_2) = \int^{y^2_1}_0 \int^{y^2_2}_0 da db
\frac{d}{da}\frac{d}{db}I(a,b)+ I(0,y^2_2)+I(y^2_1,0)-I(0,0) \label{intrick}
\end{equation}
The last three terms are of standard form and can be calculated following the
procedures of reference\cite{dolan}.  Using the identity (3.11) of reference
\cite{dolan} we find
\begin{eqnarray}
\frac{d}{da}\frac{d}{db} I(a,b) & = &
-\frac{1}{2}\frac{d}{da}\frac{d}{db}\int^{\Lambda}_0 x dx \sqrt{(x^2+a)(x^2+b)}
+ \nonumber \\ & & \int^{\Lambda}_0 x^2 dx \sum_{l=1}^{\infty} \left[\frac{2\pi
l x}{(x^2+a)(x^2+b)+(2\pi l x)}\right]^2 \label{formula}
\end{eqnarray}
where we introduced an upper momentum cutoff to regulate the individual terms.
The integral in the first term is the same as the plasma contribution to the
zero temperature effective potential in eq. (\ref{gauginveffpot}) but in terms
of the variables $a, b$.  This integral has a logarithmic cutoff
dependence. The second term does not have infrared divergences for $a, b =0$
and can be expanded in power series of $a, b$. In this series, the first term,
with $a=b=0$ has a logarithmic cutoff dependence that exactly cancels (upon
integration on $a, b$) that of the first term. The remaining series is both
ultraviolet and infrared safe. Finally only the finite terms from the {\em zero
temperature} effective potential but in terms of the variables $y^2_1; \;
y^2_2$ contribute to $ I(y^2_1,y^2_2)$ and we obtain the result quoted in
equation (\ref{highTveff}).

\acknowledgements

D.Boyanovsky would like to thank H. J. de Vega and G. Amelino-Camelia for
useful references and illuminating discussions, and NSF for support through
grant award No: PHY-9302534.

D.-S. Lee was supported in part by DOE grant DE-FG05-85ER-40219-TASKA

D.Brahm and R.Holman were partially supported by the U.S. Dept. of Energy under
Contract DE-FG02-91-ER40682.

\end{document}